\documentclass[aps,pra,nobibnotes,nofootinbib,showpacs,reprint]{revtex4-1}

\usepackage{graphicx}
\usepackage{amsmath}

\newcommand{\vect}[1]{\mathbf{#1}}

\newcommand{\ket}[1]{|#1\rangle}

\newcommand{\Fig}[1]{FIG.~\ref{#1}}
\newcommand{\Eq}[1]{Eq.~(\ref{#1})}

\newcommand{\wo}{\omega_{0}}

\newcommand{\AL}{A_{1}}
\newcommand{\AR}{A_{2}}
\newcommand{\kL}{k_{1}}
\newcommand{\kR}{k_{2}}
\newcommand{\wL}{\omega_{1}}
\newcommand{\wR}{\omega_{2}}

\begin{document}

\title{Testing Quantum Coherence in Stochastic Electrodynamics with Squeezed Schr\"{o}dinger Cat States}

\author{Wayne~Cheng-Wei~Huang}
\affiliation{Center for Fundamental Physics, Northwestern University, Evanston, Illinois 60208, USA }

\author{Herman~Batelaan}
\email{email: hbatelaan2@unl.edu} 
\affiliation{Department of Physics and Astronomy, University of Nebraska-Lincoln, Lincoln, Nebraska 68588, USA }

\begin{abstract}

The interference pattern in electron double-slit diffraction is a hallmark of quantum mechanics. A long standing question for stochastic electrodynamics (SED) is whether or not it is capable of reproducing such effects, as interference is a manifestation of quantum coherence. In this study, we use excited harmonic oscillators to directly test this quantum feature in SED. We use two counter-propagating dichromatic laser pulses to promote a ground-state harmonic oscillator to a squeezed Schr\"{o}dinger cat state. Upon recombination of the two well-separated wavepackets, an interference pattern emerges in the quantum probability distribution but is absent in the SED probability distribution. We thus give a counterexample that rejects SED as a valid alternative to quantum mechanics.

\end{abstract}


\maketitle

\section{Introduction}

Over the past decades, there has been sustained interest in developing classical alternatives to quantum mechanics (QM) with the goal of solving the quantum-classical boundary problem. Despite the proposed classical alternatives \cite{Pena1996, Pena2015, Cavalleri}, there is a lack of quantitative tests of such theories against QM, mostly because analytic solutions to concrete physical systems such as two-level atoms have not be found. Arguably one of the most developed classical alternatives is stochastic electrodynamics (SED) \cite{Milonni, Boyer1975-1}. Studies of SED harmonic systems have found many examples that are in exact agreement with QM. These include the retarded van der Waals force \cite{Boyer1973}, ground state distribution of harmonic oscillators \cite{Boyer1975-2, Huang2013}, Landau diamagnetism \cite{Boyer1980, Boyer2016}, Planck spectrum of blackbody radiation \cite{Boyer1984, Boyer2017}, and Debye specific-heat law for solids \cite{Santos}. Numerical studies of hydrogen have given some qualitative features \cite{Puthoff, Cole}, but have not led to a clear success \cite{Nieuwenhuizen2015, Nieuwenhuizen2016}. Recently, it was further shown that parametric interaction can give rise to discrete SED excitation spectra that is in excellent agreement with QM predictions \cite{Huang2015}. 

\begin{figure}[b]
\centering
\scalebox{0.3}{\includegraphics{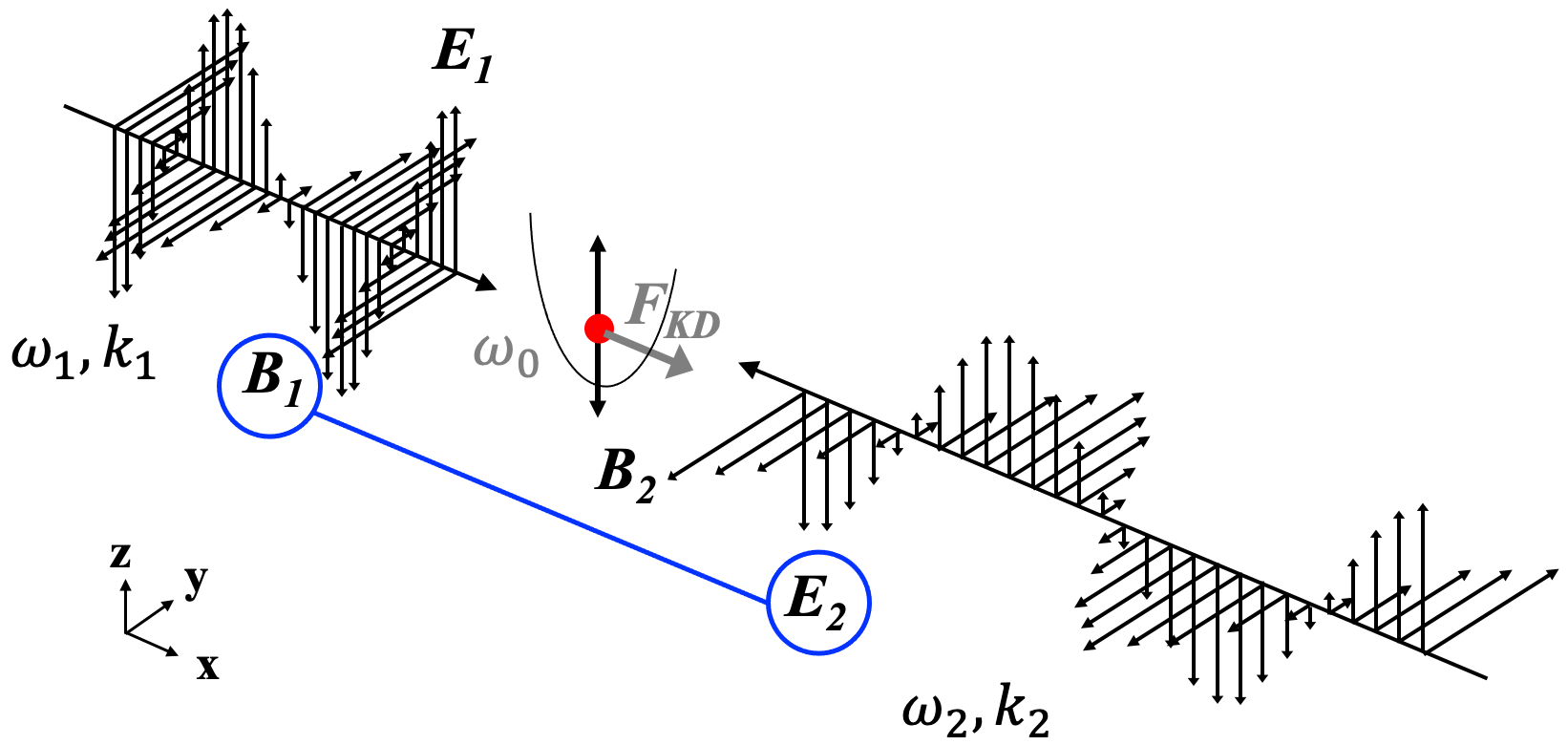}}
\caption{Two counter-propagating laser fields with frequencies $\wL$ and $\wR$, collaboratively drive the harmonic oscillator with a spatially modulated Kapitza-Dirac force in the direction of wave propagation. The force has a modulation periodicity of $2\pi c/(\wL+\wR)$ and it oscillates at the difference frequency $\wL-\wR$. A particle subject to the perturbation of the classical zero-point electromagnetic field can get pushed to either directions.}
\label{fig:KD}
\end{figure}

\begin{figure*}[t]
\centering
\scalebox{0.45}{\includegraphics{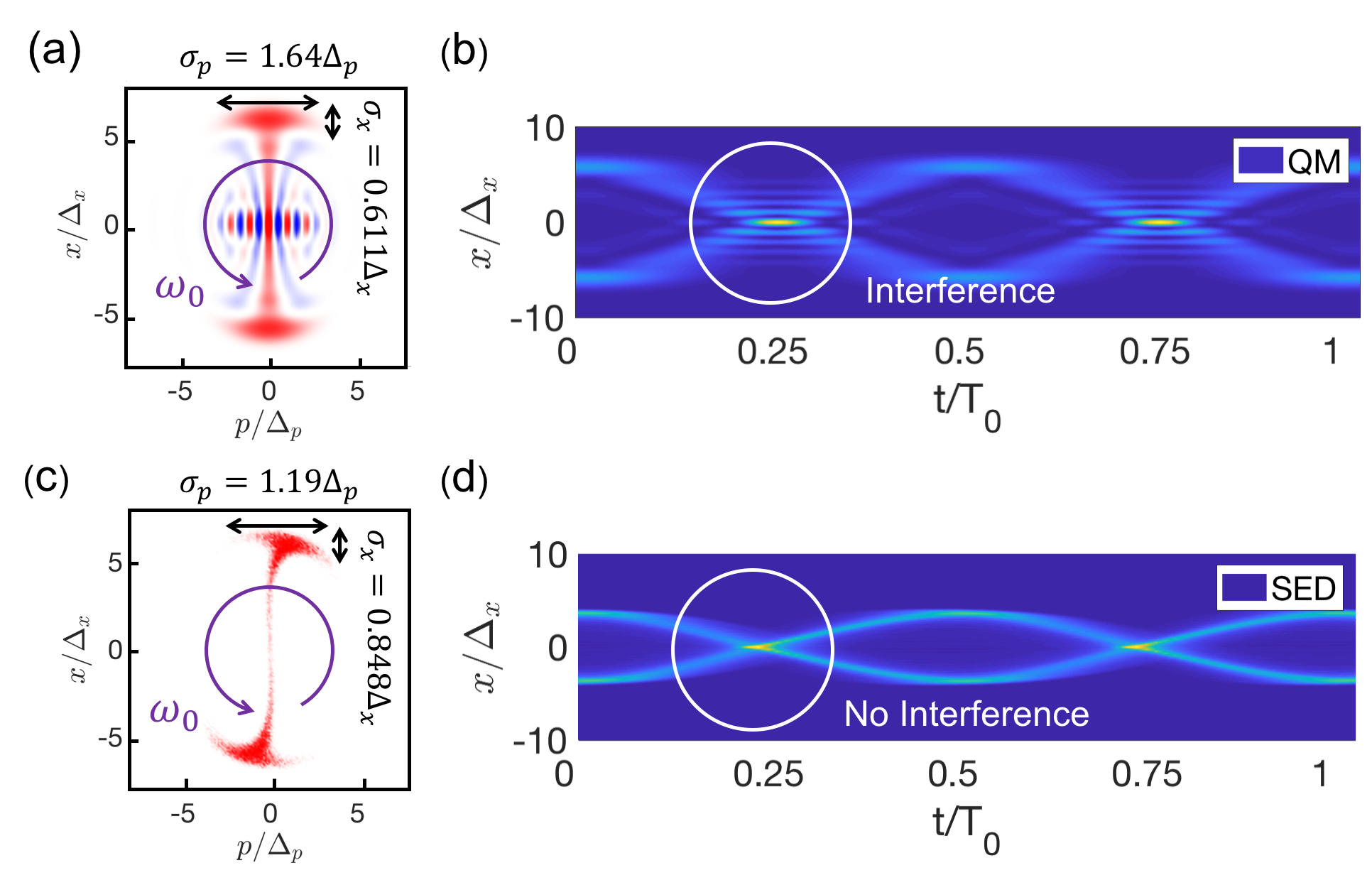}}
\caption{Time evolution of QM and SED probability distributions after laser excitation. (a) The Wigner function of the QM oscillator is plotted after the laser excitation. Positive values are color coded in red and negative values are in blue. Quantum coherence between the two well-separated squeezed states is manifested by the fringe structure in the center. The Wigner function rotates counter-clockwise at the oscillator's resonant frequency $\wo = 10^{16}$ (rad/s). At the moment when the distribution is depicted, the position quadrature uncertainty $\sigma_{x}$ of the squeezed state is smaller than that of the ground state $\Delta_{x} = \sqrt{\hbar/2m\wo}$. Meanwhile, the quadrature uncertainty product satisfies the Heisenberg relation $\sigma_{x}\sigma_{p} = \hbar/2$ at all times. (b) The two wavepackets oscillate back and forth in the harmonic potential giving rise to a double sinusoidal trajectory of the QM probability distribution. An interference pattern appears when the two wavepackets merge. (c) The phase space distribution of the SED oscillator shows two well-separated sub-ensembles. Each sub-ensemble has a squeezed structure that mimics the QM squeezed state shown in (a). (d) As the SED phase space distribution rotates counter-clockwise, the probability distribution bundles into two "macroscopic" trajectories. In each trial of the SED simulation, there is no knowledge which macroscopic trajectory a particle will follow unless the initial phase of the background zero-point field is known. No interference-like patterns are found in the SED probability distribution when the two macroscopic trajectories of distributions cross.}
\label{fig:traj}
\end{figure*}

However, a major drawback to the generality of the SED approach is that none of the investigated effects involves quantum coherence. In light of this, some have proposed to study electron double-slit diffraction within the framework of SED as interference is a manifestation of quantum coherence \cite{Boyer1975-1, Huang2013, Bach}. Within the SED community the proposed view of electron diffraction is that the double-slit poses boundary conditions that modify the classical zero-point electromagnetic field and in turn it acts as a guiding wave for free electrons \cite{Cavalleri, Boyer1975-1, Avendano, Kracklauer1992}. The appeal of this idea is that the guiding field can be affected by both slits, while the particle passes through only one slit, similar to the idea that has pushed oil droplet analogues \cite{Wolchover, Couder, Andersen, Bohr, Batelaan2016, Pucci}. As appealing as the idea may sound, so far there has been no concrete calculation or simulation demonstrating such an effect because of two major theoretical obstacles: (1) the effective spectrum of the zero-point field is unbounded for free electrons, (2) The radiation damping of free electrons gives rise to runaway solutions.

Rather than focusing on the specific theoretical difficulties that are relevant to free electrons, we develop a paradigm that can be used as a direct test for quantum coherence in stochastic electrodynamics. Building on our previous results \cite{Huang2013, Huang2015}, we devise a laser excitation scheme to promote a ground-state quantum harmonic oscillator to a squeezed Schr\"{o}dinger cat state \cite{Kienzler, Etesse, Huang}. The Schr\"{o}dinger cat state of a harmonic oscillator is an analogy to the electron double-slit state, $\ket{L}+\ket{R}$, where $\ket{L}$ and $\ket{R}$ indicate the left and right electron slit states in the position space \cite{Bach, Tonomura}. Comparing the QM probability distribution with that of SED harmonic oscillators, we observe some interesting similarities. Nevertheless, the interference pattern is missing in the SED probability distribution. 

\section{Kapitza-Dirac force on harmonic oscillators}

Let us consider the setup in \Fig{fig:KD}. Two counter-propagating laser fields propagate along the $x$-axis, and the electric fields are linearly polarized along the $z$-axis, 
\begin{equation}\label{laserfield}
	\begin{split}
		\vect{E}_{1} =& \AL\wL\cos{(\kL x-\wL t)} \hat{\epsilon}_{z}	  \\
		\vect{E}_{2} =& -\AR\wR\cos{(\kR x+\wR t)} \hat{\epsilon}_{z}	,
	\end{split}
\end{equation}
where $\hat{\epsilon}_{z}$ is the unit vector along the $z$-axis, $k_{1,2} = \omega_{1,2}/c$ are the wave numbers, and $A_{1,2}$ are amplitudes of the corresponding vector potentials $\vect{A}_{1} = \AL\sin{(\kL x-\wL t)} \hat{\epsilon}_{z}$ and $\vect{A}_{2} = \AR\sin{(\kR x+\wR t)} \hat{\epsilon}_{z}$. The electric and magnetic components of the combined laser field are
\begin{equation}\label{field}
	\begin{split}
		E_{z} = \AL\wL\cos{(\kL x-\wL t)} - \AR\wR\cos{(\kR x+\wR t)} 	\\
		B_{y} = - \AL\kL\cos{(\kL x-\wL t)} - \AR\kR\cos{(\kR x+\wR t)} 	.
	\end{split}
\end{equation}
Assuming that the particle is free in the direction of the electric field (i.e. the $z$-axis in \Fig{fig:KD}), the cross terms between the electric and magnetic components can give rise to a spatially modulated force in the direction of wave propagation (i.e. the $x$-axis in \Fig{fig:KD}). Herein we term this force the Kapitza-Dirac (KD) force. The KD force can be derived using the following equations of motion,
\begin{equation}\label{KDmotion}
	\left\{
	\begin{array}{l}
		\displaystyle	m\frac{dv_{z}}{dt} = qE_{z}	\\	
		\\	
		\displaystyle	F^{(\textrm{KD})}_{x} = qv_{z}B_{y} 
	\end{array}	
	\right.	, 
\end{equation}
where $m$ and $q$ are mass and charge of the particle. Now, there are two scenarios: (1) if frequencies of the two laser fields are identical ($\wL = \wR$), the KD force will be constant in time, which is also known as the pondermotive force \cite{Batelaan2007}, (2) if the laser frequencies are different ($\wL \ne \wR$), the KD force will oscillate in time with the sum and difference frequencies, $\wL+\wR$ and $\wL-\wR$. If the charge particle is confined by a harmonic potential $U(x) = m\wo^2x^2/2$ with a resonant frequency $\wo = (\wL-\wR)/2$, the KD force that can resonantly drive the harmonic oscillator will be (see derivation in Appendix)
\begin{equation}\label{force}
		F_{\textrm{KD}} = \frac{q^2\AL\AR}{m} \left(\frac{\kL+\kR}{2}\right)  \sin{\left( (\kL+\kR)x \right)}	\cos{\left( (\wL-\wR)t \right)}. 
\end{equation}
Accordingly, the corresponding time-varying KD potential is 
\begin{equation}\label{KDpotential}
	        U_{\textrm{KD}} = \frac{q^2\AL\AR}{2m} \cos{\left( (\kL+\kR)x \right)}\cos{\left( (\wL-\wR)t \right)}. 
\end{equation}
We note that at any given time $t = t_{0}$ a trapping site in the KD potential $U_{\textrm{KD}}(x,t_{0})$ (i.e. a minimum in the potential) will turn into an unstable point after a quarter of the natural period $T_{0}/4 = \pi/|\wL-\wR|$, where $T_{0} = 2\pi/\wo$, since the potential polarity is reversed,
\begin{equation}
	 U_{\textrm{KD}}(x,t_{0}+T_{0}/4) = -U_{\textrm{KD}}(x,t_{0}).
\end{equation}
This feature will later be used to coherently split the ground-state wavepacket of a quantum oscillator.

\section{Generation of squeezed Schr\"{o}dinger cat states}

The KD effect for quantum harmonic oscillators can be modeled by adding the KD potential in \Eq{KDpotential} to the unperturbed oscillator Hamiltonian. We replace the continuous-wave laser fields in \Eq{laserfield} with pulsed fields,
\begin{equation}\label{pulse}
	\begin{split}
	\vect{E}_{1}(x,t) &= \AL\wL\cos{(\kL x-\wL t)}e^{-(t/\tau)^2} {\hat{\epsilon}_{z}}		\\
	\vect{E}_{2}(x,t) &= -\AR\wR\cos{(\kR x+\wR t)}e^{-(t/\tau)^2}{\hat{\epsilon}_{z}}		,	
	\end{split}
\end{equation}
where $\tau$ is the pulse duration, in order to avoid indefinite sequential excitation of the oscillator's ladder levels. Given the appropriate pulse amplitudes and durations, the final population distribution can have a peaked structure. The quantum Hamiltonian is thus
\begin{equation}\label{hamiltonian}
	\hat{H} = \frac{m\wo}{2}\hat{x}^2 + \frac{\hat{p}^2}{2m} + U_{KD}(\hat{x},t)e^{-2(t/\tau)^2}.
\end{equation}
The difference frequency of the laser fields is twice the oscillator's resonant frequency $\wL-\wR = 2\wo$, so the ground-state oscillator will be parametrically excited to the even-symmetry states $\ket{n = 2k}$, which is a prerequisite for cat state generation because a cat state has an even symmetry. We obtained the QM result by numerically solving the Sch\"{o}dinger equation as in \cite{Huang2015}. The oscillator's parameters are $m = 9.11 \times 10^{-35}$ (kg), $q =1.60 \times 10^{-19}$ (C), and $\wo = 10^{16}$ (rad/s). The laser parameters are chosen to be $\wL = 2.3\, \wo$, $\wR = 0.3\, \wo$, $\tau = 5 \times 10^{-15}$ (s), and $\AL = \AR = 4.5 \times 10^{-8}$ (V$\cdot$s/m). The mass is chosen at this unusual value in order to make the computational time for SED simulation more manageable \cite{Huang2013}. In \Fig{fig:traj}(b) the probability trajectory of the excited state is shown. Upon excitation the ground-state wavepacket is coherently split into two wavepackets, so the oscillator is in a superposition of two macroscopically distinct states. The two wavepackets oscillate back and forth in the harmonic potential. As they recombine, interference fringes emerge in the probability distribution due to the quantum coherence between the two wavepackets. The quantum coherence is readily illustrated by the fringe structure in the oscillator's Wigner function as shown in \Fig{fig:traj}(a). We note that one of the two quadrature uncertainties of each wavepacket, $\sigma_{x} \equiv \sqrt{\langle x^2 \rangle - \langle x \rangle^2}$ (or $\sigma_{p} \equiv \sqrt{\langle p^2 \rangle - \langle p \rangle^2}$), is smaller than the ground state uncertainty $\Delta_{x} = \sqrt{\hbar/2m\wo}$ (or $\Delta_{p} = \sqrt{\hbar m\wo/2}$), while their product remains the same, $\sigma_{x}\sigma_{p} = \hbar/2$, at all times (see \Fig{fig:traj}(a)). This implies that the state generated by the laser excitation is a squeezed Schr\"{o}dinger cat state (that is, a superposition of two displaced squeezed states with opposite phases) \cite{Etesse, Huang}. 

\begin{figure}[t]
\centering
\scalebox{0.25}{\includegraphics{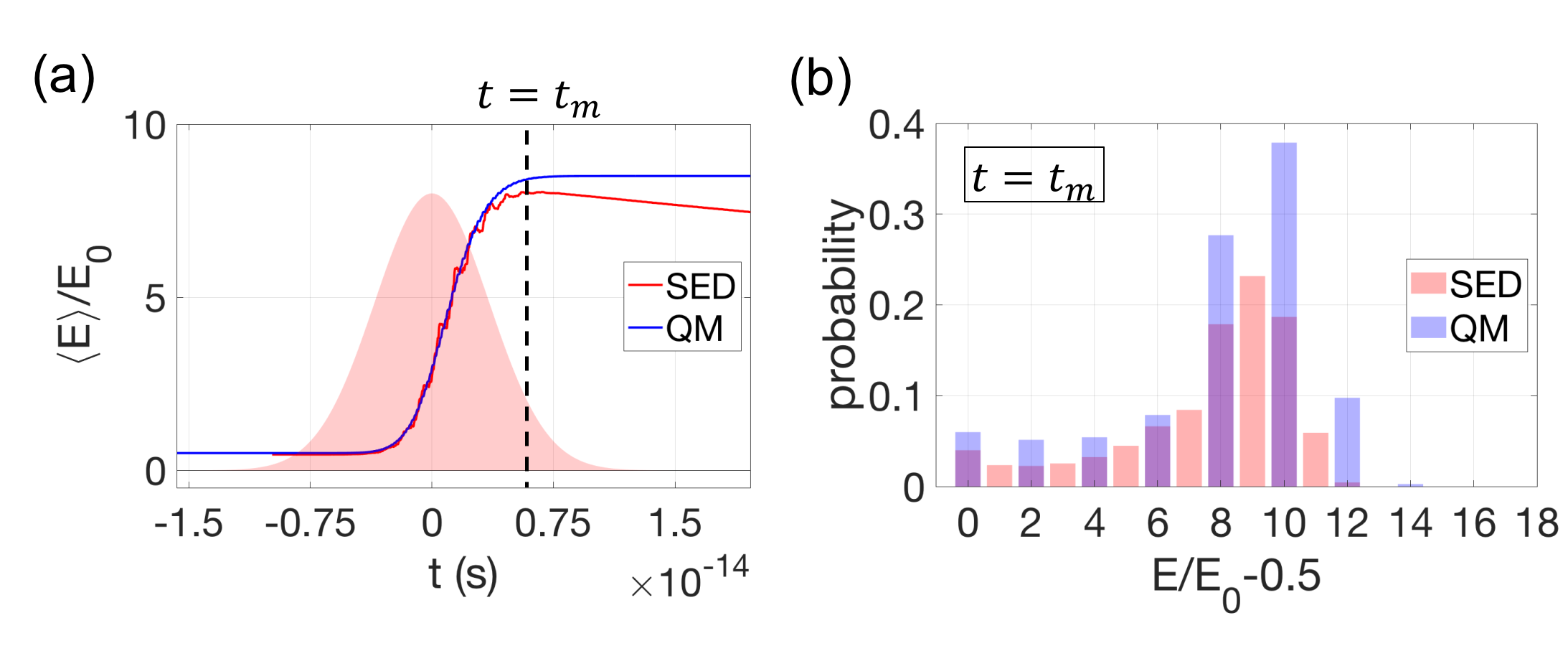}}
\caption{The QM and SED energy distributions after laser excitation. (a) The ensemble average of the SED oscillator energy (red line) is compared with the expectation value of the QM oscillator energy (blue line) during the course of laser excitation. The shaded area (red) represents the excitation laser pulse $e^{-(t/\tau)^2}$. The pulse duration is $\tau = 5 \times 10^{-15}$(s) which is much smaller than the SED oscillator damping time $\tau_{d} \approx 3.2 \times 10^{-13}$ (s). Therefore, damping has no significant effect on the oscillator's dynamics during the excitation process. (a) Energy distributions of QM and SED oscillators are compared at $t = t_{m}$ when the SED energy reaches its maximum and the radiation damping starts to dominate. The QM energy distribution (blue bar) is discrete and occupies only even energy levels $E_{2k} = E_{0}(2k+1/2)$, where $E_{0} = \hbar\wo$. The SED energy distribution (red bar) is continuous but has a similar width and average value as the QM distribution. This indicates that the excitation process is identical for QM and SED oscillators.}
\label{fig:eng}
\end{figure}

For SED oscillators, we first prepare the classical ensemble in a ground state with the $x$ and $p$ probability distributions identical to those of a quantum oscillator (see details in \cite{Huang2013}). Under excitation of the same laser pulses, the equation of motion for the SED harmonic oscillator is 
\begin{equation}\label{motion}
	m\frac{d^2x}{dt^2}= -m\wo^2x - m\Gamma\wo^2\frac{dx}{dt} + qE^{(x)}_{vac}(x, t) + F_{\textrm{KD}}(x,t)e^{-2(t/\tau)^2}	, 
\end{equation}
where $\displaystyle \Gamma \equiv \frac{2q^2}{3mc^3}\frac{1}{4\pi\epsilon_{0}}$ is the radiation damping coefficient. The zero-point field $E^{(x)}_{vac}(x, t)$ is configured according to \cite{Huang2013}. Apart from the damping term $m\Gamma\wo^2dx/dt$ and the coupling to the zero-point field $E^{(x)}_{vac}(x, t)$, \Eq{motion} is formally equivalent to a quantum Heisenberg equation derived from the Hamiltonian in \Eq{hamiltonian}. This suggests that the excitation dynamics in SED should be identical to QM assuming (1) the pulse duration $\tau$ is much shorter than the damping time $\tau_{d} = 2/\Gamma\wo^2$, and (2) the KD force is much stronger than the fluctuating force from the zero-point field \cite{Huang2013}, 
\begin{equation}
 	\frac{q^2\AL\AR}{m} \left(\frac{\kL+\kR}{2}\right)  \gg \frac{q}{2\pi}\sqrt{\frac{\hbar\Gamma\wo^5}{\epsilon_{0} c^3}}.	
\end{equation}
In our simulation these two conditions are satisfied, and the excitation dynamics in SED and QM are the same as shown in \Fig{fig:eng}(a), where the time evolutions of the expectation value of the QM energy and the ensemble average of the SED energy are compared. The two energy trajectories stay overlapped through most of the pulse and only deviate at the end of the excitation. Furthermore, the SED and QM energy distributions have similar shapes (see \Fig{fig:eng}(b)), despite that the QM distribution is discrete and the SED distribution is continuous. 

\begin{figure}[t]
\centering
\scalebox{0.35}{\includegraphics{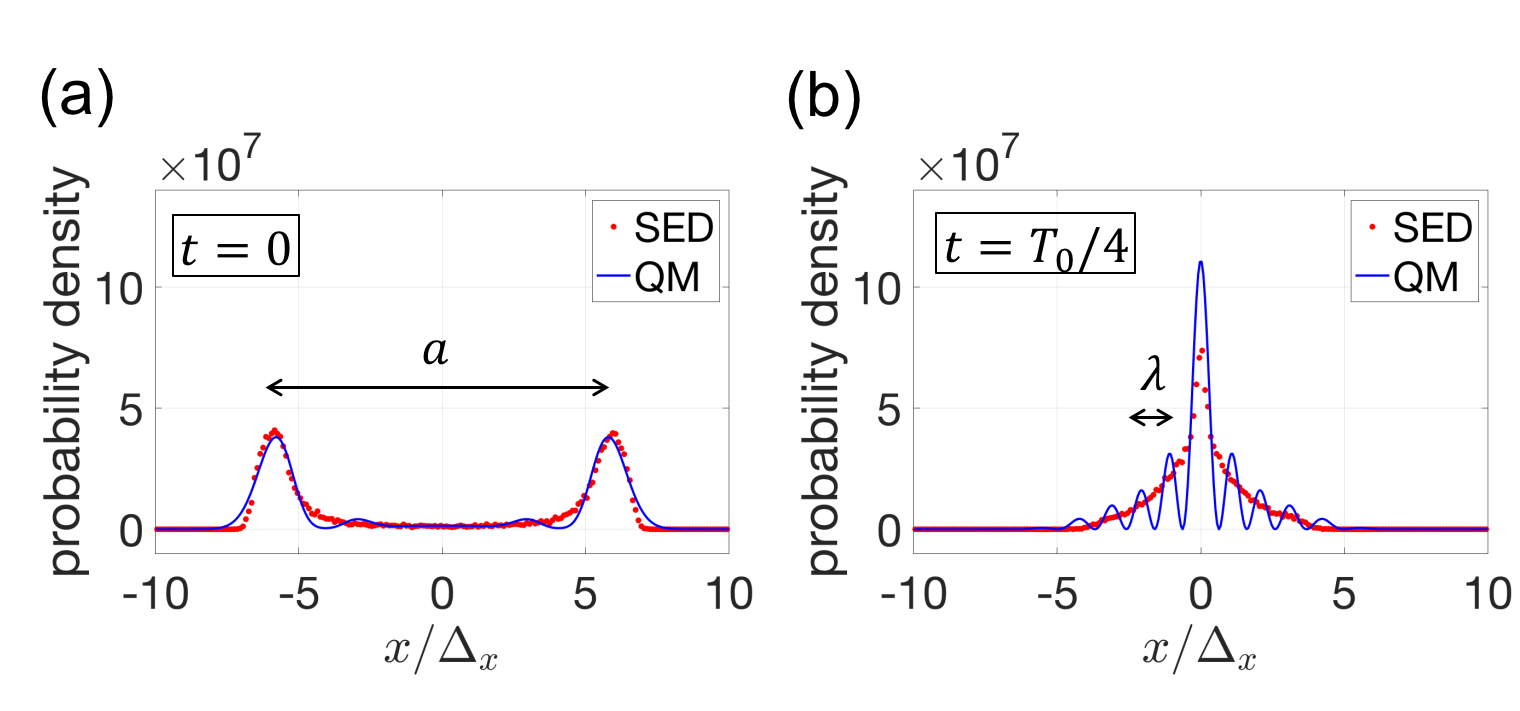}}
\caption{Comparison between the QM and SED probability distributions at $t=0$ and $t= T_{0}/4$ in \Fig{fig:traj}. (a) The agreement between QM and SED probability distributions is good when the two macroscopically distinct QM wavepackets are well-separated by a peak-to-peak distance $a$ beyond the ground state width $\Delta_{x}$, $a \gg \Delta_{x}$. (b) After a quarter of the natural period $T_{0}/4$, the two QM wavepackets recombine in the harmonic potential. Interference fringes appear in the QM distribution (blue line) but not in the SED distribution (red dots). The fringe periodicity $\lambda$ is determined by the wavepacket separation $a$ in (a) through the relation $\lambda = 2\pi\hbar/m\wo a$, which resembles the well-known double-slit diffraction formula. The SED probability distribution captures the outline of the QM distribution as if the quantum coherence between the two QM wavepackets is lost.}
\label{fig:prob}
\end{figure}

The probability trajectory of the SED oscillator ensemble along with its phase space distribution are shown in \Fig{fig:traj}(c)(d). The ensemble particle number is $N_{p} = 3 \times 10^4$. The parametric interaction between the SED oscillator and the laser fields were simulated using the same method as in \cite{Huang2015}. Like the QM oscillator, upon excitation the ground-state SED probability distribution also splits into two sub-ensembles that follow two distinct sinusoidal trajectories. The initial phase spectrum of the zero-point field, which is random and considered as the``hidden variable", determines which trajectory a particle will follow in each trial. A detailed comparison between SED and QM probability distributions is given in \Fig{fig:prob} for two time points: (1) when the two QM wavepackets are separated, (2) when they recombine. Although there are some overall similarities between the SED and QM distributions, there are no interference fringes in the SED probability distribution.

\section{Discussion and Conclusion}

While the zero-point electromagnetic field only introduces small radiative corrections to nonrelativistic QM, such as Lamb shifts, it drastically changes the particle dynamics in classical mechanics and leads to the reproduction of QM effects in some classical systems \cite{Boyer1973,Boyer1975-2, Huang2013,Boyer1980,Boyer1984,Boyer2017,Santos,Huang2015}. Our work aims to investigate to what extent such a classical theory can reproduce QM features by comparing results obtained from SED, \Eq{motion}, with those obtained from the QM Hamiltonian, \Eq{hamiltonian}. The qualitative difference between the probability distributions of QM and SED in \Fig{fig:traj} and \Fig{fig:prob}(b) establishes that SED in its traditional form does not support physical effects that involve quantum coherence \cite{Milonni, Boyer1975-1}. The squeezed Schr\"{o}dinger cat state used in this work is an analogy to the electron double-slit experiment \cite{Bach}. The peak-to-peak separation $a$ between the two wavepackets in \Fig{fig:prob}(a) determines the fringe periodicity $\lambda$ in \Fig{fig:prob}(b) through the relation $\lambda = 2\pi\hbar/m\wo a$, which mimics the double-slit diffraction formula. Our analysis provides evidence that coherence-like behavior is absent in SED, and thus we predict that SED electron double-slit diffraction, if ever calculated, will not show fringes.

On the other hand, our result helps to establish the validity range of SED. We note that the partial agreement between the SED and QM results stems from the formal resemblance between the SED equation of motion and the QM Heisenberg equation assuming that the laser excitation pulses satisfy certain criteria. While $x$ and $p$ are independent dynamic variables in SED, the canonical commutation relation $[\hat{x},\hat{p}] = i\hbar$ makes them a Fourier pair in QM. This difference makes the distinction between QM and SED in terms of quantum coherence. Therefore, SED may be seen as the decoherence limit of QM \cite{Sonnentag}. Although the zero-point field bestows a special phase relation between $x$ and $p$ of a SED harmonic oscillator, which leads to the quantum ground-state distributions \cite{Boyer1975-2, Huang2013}, the phase relation serves only as initial conditions and does not affect the dynamical evolution of $x$ and $p$ during laser excitation. Therefore, we speculate that any mechanism or theoretical operation that restores (or deteriorates) the Fourier relation between $x$ and $p$ for SED (or QM), will make the proper decoherence theory that bridges the gap between SED and QM. 

\section{Acknowledgement}

The authors thank A. M. Steinberg, C. Monroe, and P. W. Milonni, for advice. W. C. Huang wishes to give special thanks to Yanshuo Li for helpful discussions. This work utilized high-performance computing resources from the Holland Computing Center of the University of Nebraska. We gratefully acknowledge funding support from NSF PHY-1602755. 

\appendix*

\section{Derivation of the resonant Kapitza-Dirac force}\label{app}

In this appendix we derive the KD force in \Eq{force} from \Eq{KDmotion} using the electric and magnetic components of the combined laser field given in \Eq{field}. First, we solve the velocity $v_{z}$ by integrating the equation of motion $mdv_{z}/dt = qE_{z}$, 
\begin{equation}
	v_{z}(t) = - \frac{q}{m}\left( \AL\sin{(\kL x-\wL t)} + \AR\sin{(\kR x+\wR t)} \right). 
\end{equation}
Substituting $v_{z}(t)$ to $F_{x} = qv_{z}B_{y}$, we obtain the Lorentz force
\begin{equation}\label{lorentz}
	\begin{split}
	F_{x} =& \frac{q^2}{m}\left( \AL\sin{(\kL x-\wL t)} + \AR\sin{(\kR x+\wR t)} \right)	\\
		    & \times \left(	 \AL\kL\cos{(\kL x-\wL t)} + \AR\kR\cos{(\kR x+\wR t)} \right). 
	\end{split}
\end{equation}
We can see four frequency components in the Lorentz force by expanding \Eq{lorentz},
\begin{widetext}
\begin{equation}
	\begin{split}
	F_{x} =& \frac{q^2}{2m}\left[ \AL^2\kL\sin{(2\kL x-2\wL t)} \right.	\\
		   & + \AL\AR\kL \left( \sin{((\kL+\kR) x-(\wL-\wR)t)}-\sin{((\kL-\kR) x-(\wL+\wR)t)} \right)	\\
		   & + \AL\AR\kR \left( \sin{((\kL+\kR) x-(\wL-\wR)t)}+\sin{((\kL-\kR) x-(\wL+\wR)t)} \right)	\\
		   & + \left.	\AL^2\kR\sin{(2\kR x+2\wR t)}	\right]	.
	\end{split}
\end{equation}
\end{widetext}
Because the frequency components $2\wL$, $2\wR$, and $\wL+\wR$ are not integer multipliers of $\wo$, we can drop these terms and keep only the parametrically resonant term $\wL-\wR=2\wo$, 
\begin{equation}
	F_{x} \approx \frac{q^2}{2m} \AL\AR (\kL+\kR)  \sin{((\kL+\kR) x-(\wL-\wR)t)}	.
\end{equation}
The force has a traveling wave profile which can be decomposed into two standing-wave components with even and odd symmetries,
\begin{widetext}
\begin{equation}
	\sin{((\kL+\kR) x-(\wL-\wR)t)} = \sin{((\kL+\kR) x)}\cos{((\wL-\wR)t)} -\cos{((\kL+\kR) x)}\sin{((\wL-\wR)t)}.
\end{equation}
\end{widetext}
The force needs to have a potential profile with even symmetry in order to resonantly drive the oscillator from the ground state with an even frequency ($\wL-\wR=2\wo$). This implies that the resonant KD force should an odd symmetry, thus it takes the form
\begin{equation}
	F_{KD} = \frac{q^2}{2m} \AL\AR (\kL+\kR) \sin{((\kL+\kR) x)}\cos{((\wL-\wR)t)}	.
\end{equation}


\end{document}